\documentclass{article}
\def\Ref#1{(\ref{#1})}
\usepackage{amsmath}
\usepackage{amssymb}
\usepackage{cite}
\def\d{{\rm d}}
\def\bo{{\bigcirc}}

\begin{document}
\begin{titlepage}
\noindent{\large\textbf{ Exactly solvable  reaction diffusion
models on a Cayley tree}}

\vskip 2 cm

\begin{center}{Laleh~Farhang Matin{\footnote
{laleh.matin@alzahra.ac.ir}}, Amir~Aghamohammadi{\footnote
{mohamadi@alzahra.ac.ir}}, \& Mohammad~Khorrami{\footnote
{mamwad@mailaps.org}} } \vskip 5 mm \textit{ Department of
Physics, Alzahra University,
             Tehran 1993891167, Iran. }

\end{center}

\begin{abstract}
\noindent The most general reaction-diffusion model on a Cayley
tree with nearest-neighbor interactions is introduced, which can
be solved exactly through the empty-interval method. The
stationary solutions of such models, as well as their dynamics,
are discussed. Concerning the dynamics, the spectrum of the
evolution Hamiltonian is found and shown to be discrete, hence
there is a finite relaxation time in the evolution of the system
towards its stationary state.
\end{abstract}
\begin{center} {\textbf{PACS numbers:}} 05.40.-a, 02.50.Ga

 \end{center}
\end{titlepage}
\section{Introduction}
Reaction-diffusion systems have been studied using various
methods, including analytical techniques, approximation methods,
and simulation. Approximation methods are generally different in
different dimensions, as for example the mean field techniques,
working good for high dimensions, generally do not give correct
results for low-dimensional systems. A large fraction of
analytical studies belong to low-dimensional (specially
one-dimensional) systems, as solving low-dimensional systems
should in principle be easier.
\cite{ScR,ADHR,KPWH,HS1,PCG,HOS1,HOS2,AL,AKK,AKK2,AM1}.

The Cayley tree is a tree (a lattice having no loops) where every
site is connected to $\xi$ nearest neighbor sites. This no-loops
property may allow exact solvability for some models, for general
coordination number $\xi$. Reaction diffusion models on the Cayley
tree have been studied in, for example
\cite{VNH,jK,SNPVP,MKMS,SNM,ACVP}. In \cite{VNH,jK,SNM}
diffusion-limited aggregations, and in \cite{SNPVP} two-particle
annihilation reactions for immobile reactants have been studied.
There are also some exact results for deposition processes on the
Bethe lattice \cite{ACVP}.

The empty interval method (EIM) has been used to analyze the one
dimensional dynamics of diffusion-limited coalescence
\cite{BDb,BDb1,BDb2,BDb3}. Using this method, the probability that
$n$ consecutive sites are empty has been calculated. This method
has been used to study a reaction-diffusion process with
three-site interactions \cite{HH}. EIM has been also generalized
to study the kinetics of the $q$-state one-dimensional Potts model
in the zero-temperature limit \cite{MB}. In
\cite{BDb,BDb1,BDb2,BDb3}, one-dimensional diffusion-limited
processes have been studied using EIM. There, some of the reaction
rates have been taken infinite, and the models have been worked
out on continuum. For the cases of finite reaction-rates, some
approximate solutions have been obtained.

In \cite{AKA,AM3}, all the one dimensional reaction-diffusion
models with nearest neighbor interactions which can be exactly
solved by EIM have been found and studied. Conditions have been
obtained for the systems with finite reaction rates to be solvable
via EIM, and then the equations of EIM have been solved. In
\cite{AKA}, general conditions were obtained for a single-species
reaction-diffusion system with nearest neighbor interactions, to
be solvable through EIM. Here solvability means that evolution
equation for $E_n$ (the probability that $n$ consecutive sites be
empty) is closed. It turned out there, that certain relations
between the reaction rates are needed, so that the system is
solvable via EIM. The evolution equation of $E_n$ is a recursive
equation in terms of $n$, and is linear. It was shown that if
certain reactions are absent, namely reactions that produce
particles in two adjacent empty sites, the coefficients of the
empty intervals in the evolution equation of the empty intervals
are $n$-independent, so that the evolution equation can be easily
solved. The criteria for solvability, and the solution of the
empty-interval equation were generalized to cases of multi-species
systems and multi-site interactions in \cite{KAA,AAK,AK}.

In this article  the most general single-species
reaction-diffusion model with nearest-neighbor interactions on a
Cayley tree is investigated, which can be solved exactly through
the empty interval method. The scheme of the paper is as follows.
In section 2, the most general reaction-diffusion model with
nearest-neighbor interactions on a Cayley tree is studied, which
can be solved exactly through EIM. The evolution equation of $E_n$
is also obtained. In section 3 the stationary solution of such
models, as well as their dynamics are discussed. Finally, section
4 is devoted to concluding remarks.
\begin{figure}
\begin{picture}(170,170)
\includegraphics{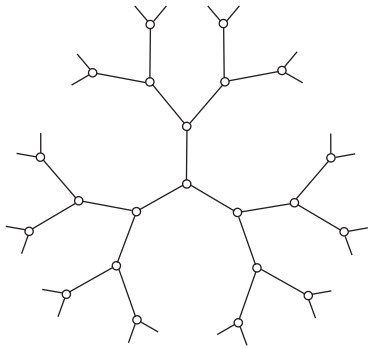}
\end{picture}
\caption{The Cayley tree with $\xi =3$ }
\end{figure}
\begin{figure}
\begin{picture}(170,170)
\includegraphics{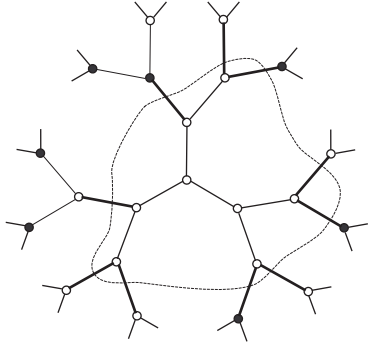}
\end{picture}
\caption{An empty cluster with the links at the boundary, on a
Cayley tree with $\xi =3$ }
\end{figure}
\section{Models solvable through the empty interval method on a Cayley tree}
The Cayley tree is a tree (a lattice without loops) where each
site is connected to $\xi$ sites (fig. 1). Two sites are called
neighbors iff they are connected through a link. Consider a system
of particles on a Cayley tree. Each site is either empty or
occupied by one particle. The interaction (of particles and
vacancies) is nearest neighbor. The probability that a connected
collection of $n$ sites be empty is denoted by $E_n$. It is
assumed that this quantity does not depend on the choice of the
collection. An example is a tree where the probability that a site
is occupied is $\rho$ and is independent of the states of other
sites. Then
\begin{equation}\label{ca.01}
E_n=(1-\rho)^n.
\end{equation}
The following graphical representations help express various
relations in a more compact form. An empty (occupied) site is
denoted by $\circ$ ($\bullet$). A connected collection of $n$
empty sites is denoted by $\bo_n$.

There is no loop in a Cayley tree, so each site can only be
connected to a single existing cluster site, by a single link. For
$\xi\geq 3$ (the case we are interested in here) the closedness of
the evolution equation for $E_n$ requires that the rate of
creating an empty site be zero. The reason is that if it is not
the case, then an empty $n$-cluster can be created from two
disjoint empty clusters joined by a single occupied site
\cite{bAG}. This shows that if the evolution of the empty clusters
is to be closed, then the only possible reactions are the
following, with the rates indicated.
\begin{align}\label{ca.02}
\bullet\circ\to&\bullet\bullet,\quad r_1\cr
\circ\circ\to&\circ\bullet,\quad r_2\cr
\circ\circ\to&\bullet\bullet,\quad r_3.
\end{align}
(There is no distinction between left and right, of course.) This
means that the reactants are immobile, and the coagulation and
diffusion rates are zero.

Using these, one arrives at the following time evolution for
$E_n$:
\begin{align}\label{ca.03}
    \frac{\d E_n}{\d t}=&-R_n\,r_1\,P(\bullet\hskip-0.14cm-
    \hskip-0.14cm\bo_n)-R_n\,(r_2+r_3)\,P(\circ\hskip-0.14cm-
    \hskip-0.14cm\bo_n)\cr & \cr &
    -(n-1)\,(2\,r_2+r_3)\,P(\bo_n),
\end{align}
where $R_n$ is the number of sites adjacent to a collection of $n$
connected sites. A simple induction shows that
\begin{equation}\label{ca.04}
R_n=n\,(\xi -2)+2.
\end{equation}
One has
\begin{equation}\label{ca.05}
P(\bullet\hskip-0.14cm- \hskip-0.14cm\bo_n)+P(\circ\hskip-0.14cm-
\hskip-0.14cm\bo_n)=P(\bo_n),
\end{equation}
from which
\begin{equation}\label{ca.06}
P(\bullet\hskip-0.14cm- \hskip-0.14cm\bo_n)=E_n-E_{n+1}.
\end{equation}
Using this, one arrives at
\begin{equation}\label{ca.07}
\frac{\d E_n}{\d t}=R_n\,[-r_1\,(E_n-E_{n+1})-(r_2+r_3)\,E_{n+1}]
-(n-1)\,(2\,r_2+r_3)\,E_n.
\end{equation}
Throughout the paper, it is assumed that $r_1$, $r_2$, and $r_3$
are all nonzero.
\section{The solution}
The stationary solution of the system ($E^{\mathrm{s}}$, for which
the time derivative vanishes), satisfies
\begin{equation}\label{ca.08}
R_n\,[-r_1\,(E^{\mathrm{s}}_n-E^{\mathrm{s}}_{n+1})-(r_2+r_3)
\,E^{\mathrm{s}}_{n+1}]-(n-1)\,(2\,r_2+r_3)\,E^{\mathrm{s}}_n=0.
\end{equation}
As $E_n$'s are nonnegative and nonincreasing in $n$, it is easy to
see that the only solution to \Ref{ca.08} is
\begin{equation}\label{ca.09}
E^{\mathrm{s}}_n=0.
\end{equation}
This means that in the stationary configuration, all of the sites
are occupied, which is not a surprise since in all reactions
particles are created.

Regarding dynamics, one question is to obtain the spectrum of the
evolution Hamiltonian. This is equivalent to finding solutions
with exponential time dependence:
\begin{equation}\label{ca.10}
E^{\mathcal{E}}_n(t)=E^{\mathcal{E}}_n\,\exp(\mathcal{E}\,t).
\end{equation}
Putting this in \Ref{ca.07}, one arrives at
\begin{equation}\label{ca.11}
-[R_n\,r_1+(n-1)\,(2\,r_2+r_3)+\mathcal{E}]\,E^{\mathcal{E}}_n
+R_n\,(r_1-r_2-r_3)\,E^{\mathcal{E}}_{n+1}=0.
\end{equation}
From this,
\begin{equation}\label{ca.12}
E^{\mathcal{E}}_{n+1}=\zeta_n\,E^{\mathcal{E}}_n,
\end{equation}
where
\begin{equation}\label{ca.13}
\zeta_n:=\frac{R_n\,r_1+(n-1)\,(2\,r_2+r_3)+\mathcal{E}}
{R_n\,(r_1-r_2-r_3)}.
\end{equation}
It is seen that
\begin{equation}\label{ca.14}
\lim_{n\to\infty}\zeta_n=\frac{(\xi-2)\,r_1+2\,r_2+r_3}{(\xi-2)\,(r_1-r_2-r_3)}.
\end{equation}
The right-hand side is either negative or greater than one. So if
all $E^{\mathcal{E}}_n$'s are nonzero, then $E^{\mathcal{E}}_n$'s
either are not all nonnegative or blow up for large $n$'s. Such
$E^{\mathcal{E}}_n$'s are not acceptable as probabilities. To see
the reason, consider $\mathcal{E}_1$ (the largest $\mathcal{E}$).
for large times, only $E_n$'s corresponding to this eigenvalue
survive. But these should be nonincreasing with respect to $n$,
and nonnegative, which is not the case. So $E^{\mathcal{E}_1}_n$'s
must be identically zero for $n$ larger than a certain integer
(say $n_1$). A similar reasoning can then be made for
$\mathcal{E}_2$ (the next largest value of $\mathcal{E}$), and the
values of $E^{\mathcal{E}_2}_n$ for $n>n_1$, to show that there
should be another integer $n_2$ so that $E^{\mathcal{E}_2}_n$
vanishes for $n>n_2$. This argument can be continued to show that
for all $\mathcal{E}$'s, there must be an integer so that
$E^{\mathcal{E}}_n$'s are identically zero for $n$ larger than
that integer. This shows that $\zeta_n$ must be zero for some
positive $n$, which gives the allowed values of $\mathcal{E}$:
\begin{equation}\label{ca.15}
\mathcal{E}_k=-\xi\,r_1-(k-1)\,\beta,\qquad k\geq 1,
\end{equation}
where
\begin{equation}\label{ca.16}
\beta:=(\xi-2)\,r_1+2\,r_2+r_3.
\end{equation}
This spectrum is discrete, and there is a gap between the largest
eigenvalue and zero, which means that the system evolves towards
its stationary configuration with a relaxation time. This
relaxation time is
\begin{equation}\label{ca.17}
\tau=\frac{1}{\xi\,r_1}.
\end{equation}
One can also find $E^{\mathcal{E}}_n$'s. Denoting
$E^{\mathcal{E}_k}_n$ by $E^k_n$, and using \Ref{ca.12} and
\Ref{ca.15}, one arrives at
\begin{equation}\label{ca.18}
E_n^k=\frac{\displaystyle{\Gamma\left(k+\frac{2}{\xi-2}\right)\,\alpha^{k-n}}}
{\displaystyle{\Gamma\left(n+\frac{2}{\xi-2}\right)\,(k-n)!}},
\end{equation}
where
\begin{equation}\label{ca.19}
\alpha:=\frac{(\xi-2)\,(r_2+r_3-r_1)}{(\xi-2)\,r_1+2\,r_2+r_3}.
\end{equation}

The general solution to \Ref{ca.07} is then
\begin{equation}\label{ca.20}
E_n(t)=\sum_{k=1}^\infty c_k\,E^k_n\,\exp(\mathcal{E}_k\,t),
\end{equation}
where $c_k$'s are to be determined from the initial condition.

A special solution to \Ref{ca.07} is of the form
\begin{equation}\label{ca.21}
E_n(t)=E_1(t)\,[b(t)]^{n-1}.
\end{equation}
Putting this in \Ref{ca.07}, one arrives at
\begin{align}\label{ca.22}
\frac{\d b}{\d t}&=-\beta\,b-\beta\,\alpha\,b^2,\cr \frac{\d
E_1}{\d t}&=-
\left(\xi\,r_1+\frac{\xi}{\xi-2}\,\alpha\,\beta\,b\right)\,E_1.
\end{align}
These are readily solved and one obtains
\begin{align}\label{ca.23}
    b(t)=&\frac{b(0)\exp(-\beta\,t)}{1+\alpha\,b(0)\,[1-\exp(-\beta\,t)]},\cr
    &\cr
    E_1(t)=&E_1(0)\exp(-\xi\,r_1\,t)\,
    \left\{\frac{1}{1+\alpha\,b(0)\,[1-\exp(-\beta\,t)]}\right\}
    ^{\frac{\xi}{\xi-2}}.
\end{align}
Using these, one obtains
\begin{equation}\label{ca.24}
E_n(t)=E_n(0)\exp[-\xi\,r_1\,t-(n-1)\,\beta\,t]\,
    \left\{\frac{1}{1+\alpha\,b(0)\,[1-\exp(-\beta\,t)]}\right\}
    ^{\frac{\xi}{\xi-2}+n-1}.
\end{equation}
It is seen that for large times, all $E_n$'s tend to zero. In fact
they decay like
\begin{equation}\label{ca.25}
E_n(t)\sim \,\exp[-\xi\,r_1\,t-(n-1)\,\beta\,t].
\end{equation}
One notes that in fact $E_n(t)$ decays like
$\exp(-\mathcal{E}_n\,t)$, and this is expected, as $E_n^k$ is
zero for $k<n$.

A special case where the ansatz \Ref{ca.21} works is the case of
initially uncorrelated-sites, so that each site is occupied with
probability $\rho$ regardless of other sites. One has then
\begin{equation}\label{ca.26}
E_n(0)=(1-\rho)^n,
\end{equation}
so that
\begin{align}\label{ca.27}
E_1(0)=&1-\rho,\cr b(0)=&1-\rho.
\end{align}

The special case $\xi=2$ can be treated directly or as a limiting
case of the general problem. The results corresponding to
\Ref{ca.15} and \Ref{ca.18} would be
\begin{equation}\label{ca.28}
\mathcal{E}_k=-2\,r_1-(k-1)\,(2\,r_2+r_3),\qquad \xi=2,
\end{equation}
and
\begin{equation}\label{ca.29}
E^k_n=\frac{1}{(k-n)!}\,\left[\frac{2\,(r_2+r_3-r_1)}{2\,r_2+r_3}\right]^{k-n},
\qquad \xi=2.
\end{equation}
Finally, the solutions corresponding to the ansatz \Ref{ca.21}
would be
\begin{equation}\label{ca.30}
b(t)=b(0)\,\exp[-(2\,r_2+r_3)\,t],\qquad \xi=2,
\end{equation}
and
\begin{align}\label{ca.31}
E_1(t)=&\
E_1(0)\,\exp\left\{\frac{2\,(r_1-r_2-r_3)}{2\,r_2+r_3}\,b(0)\,
\Big[1-\exp[-(2\,r_2+r_3)\,t]\Big]\right\}\cr &\
\times\exp(-2\,r_1\,t),\qquad \xi=2,
\end{align}
so that
\begin{align}\label{ca.32}
E_n(t)=&\
E_n(0)\,\exp\left\{\frac{2\,(r_1-r_2-r_3)}{2\,r_2+r_3}\,b(0)\,
\Big[1-\exp[-(2\,r_2+r_3)\,t]\Big]\right\}\cr &\
\times\exp\{-[2\,r_1+(n-1)\,(2\,r_2+r_3)]\,t\},\qquad \xi=2,
\end{align}
\section{Concluding remarks}
The most general single-species exclusion model on a Cayley tree
was considered, for which the evolution of the empty-intervals is
closed. It was shown that in the stationary configuration of such
models all sites are occupied. The dynamics of such systems were
also studied and it was shown that the spectrum of the evolution
Hamiltonian is discrete. The time evolution of the initially
uncorrelated system was also obtained. Among the questions
remaining, one can mention the problem of Cayley trees with
boundaries, with injection and extraction at the boundaries.
\\
\\
\textbf{Acknowledgement}: The authors would like to thank
Daniel~ben-Avraham for his very useful comments. This work was
partially supported by the research council of the Alzahra
University.


\newpage
\begin{thebibliography}{99}
\bibitem{ScR}  G. M. Sch\"{u}tz; ``Exactly solvable models for many-body
               systems far from equilibrium'' in ``Phase
               transitions and critical phenomena, vol. \textbf{19}'',
               C. Domb \& J. Lebowitz (eds.), (Academic
               Press, London, 2000).
\bibitem{ADHR} F. C. Alcaraz, M. Droz, M. Henkel, \& V. Rittenberg;
               Ann. Phys. (N.~Y.) \textbf{230} (1994) 250.
\bibitem{KPWH} K. Krebs, M. P. Pfannmuller, B. Wehefritz, \&
               H. Hinrichsen; J. Stat. Phys. \textbf{78}[FS] (1995) 1429.
\bibitem{HS1}  H. Simon; J. Phys. \textbf{A28} (1995) 6585.
\bibitem{PCG}  V. Privman, A. M. R. Cadilhe, \& M. L. Glasser; J. Stat.
               Phys. \textbf{81} (1995) 881.
\bibitem{HOS1} M. Henkel, E. Orlandini, \& G. M. Sch\"utz; J. Phys.
               \textbf{A28} (1995) 6335.
\bibitem{HOS2} M. Henkel, E. Orlandini, \& J. Santos; Ann. of Phys.
               \textbf{259} (1997) 163.
\bibitem{AL}   A. A. Lushnikov; Sov. Phys. JETP \textbf{64} (1986) 811
               [Zh. Eksp. Teor. Fiz. \textbf{91} (1986) 1376].
\bibitem{AKK}  M. Alimohammadi, V. Karimipour, \& M. Khorrami; Phys. Rev.
               \textbf{E57} (1998) 6370.
\bibitem{AKK2} M. Alimohammadi, V. Karimipour, \& M. Khorrami; J. Stat.
               Phys. \textbf{97} (1999) 373.
\bibitem{AM1}  A. Aghamohammadi \& M. Khorrami; J. Phys. \textbf{A33} (2000) 7843.
\bibitem{VNH} J. Vannimenus, B. Nickel, \& V. Hakim; Phys. Rev. \textbf{B30}
              (1984) 391.
\bibitem{jK}   J. Krug; J. Phys. \textbf{A21} (1988) 4637.
\bibitem{SNPVP} S. N. Majumdar \& V. Privman; J. Phys. \textbf{A26} (1993) L743.
\bibitem{MKMS} M. Ya. Kelbert \& Yu. M. Suhov; Comm. Math. Phys. \textbf{167} (1995)
               607.
\bibitem{SNM}  S. N. Majumdar, Phys. Rev. \textbf{E68} (2003) 026103.
\bibitem{ACVP}  A. Cadilhe \& V. Privman; Mod. Phys. Lett.
               \textbf{B18} (2004) 207 (2004).
\bibitem{BDb}  M. A. Burschka, C. R. Doering, \& D. ben-Avraham;  Phys.
               Rev. Lett. \textbf{63} (1989) 700.
\bibitem{BDb1} D. ben-Avraham;  Mod. Phys. Lett. \textbf{B9} (1995)
               895.
\bibitem{BDb2} D. ben-Avraham; in ``Nonequilibrium Statistical
               Mechanics in One Dimension'', V. Privman (ed.), pp 29-50
               (Cambridge University press,1997).
\bibitem{BDb3} D. ben-Avraham; Phys. Rev. Lett. \textbf{81} (1998)
               4756.
\bibitem{HH}   M. Henkel \& H. Hinrichsen; J. Phys. \textbf{A34}, 1561-1568
               (2001).
\bibitem{MB}   M. Mobilia \& P. A. Bares; Phys. Rev. \textbf{E64} (2001) 066123.
\bibitem{AM3}  A. Aghamohammadi \& M. Khorrami; Eur. Phys. J. \textbf{B47} (2005)
               583–586.
\bibitem{AKA}  M. Alimohammadi, M. Khorrami, \& A. Aghamohammadi;
               Phys. Rev. \textbf{E64} (2001) 056116.
\bibitem{KAA}  M. Khorrami, A. Aghamohammadi, \& M. Alimohammadi;
               J. Phys. \textbf{A36} (2003) 345.
\bibitem{AAK}  A.~Aghamohammadi, M.~Alimohammadi, \& M.~Khorrami;
               Eur. Phys. J. \textbf{B31} (2003) 371.
\bibitem{AK}   A.~Aghamohammadi \& M.~Khorrami; Int. J. Mod. Phys.
               \textbf{B18} (2004) 2047.
\bibitem{bAG}  D.~ben-Avraham \& M.~L.~Glasser; cond-mat/06120809.
\end{thebibliography}
\end{document}